Regular Article

# Controlling transient chaos in the Lorenz system with machine learning

David Valle[a], Rubén Capeans[b], Alexandre Wagemakers[c], and Miguel A. F. Sanjuán[d]

Nonlinear Dynamics, Chaos and Complex Systems Group, Departamento de Física, Universidad Rey Juan Carlos, Tulipán s/n, Móstoles, Madrid 28933, Spain

**Abstract** This paper presents a novel approach to sustain transient chaos in the Lorenz system through the estimation of safety functions using a transformer-based model. Unlike classical methods that rely on iterative computations, the proposed model directly predicts safety functions without requiring fine-tuning or extensive system knowledge. The results demonstrate that this approach effectively maintains chaotic trajectories within the desired phase space region, even in the presence of noise, making it a viable alternative to traditional methods. A detailed comparison of safety functions, safe sets, and their control performance highlights the strengths and trade-offs of the two approaches.

## 1 Introduction

Chaos theory has long captivated researchers due to its intricate dynamics and wide-ranging implications across fields such as meteorology, engineering, and biology [1]. Among its foundational models, the Lorenz system stands out as a key representation of chaotic behavior [2]. Originally developed to model atmospheric convection, the Lorenz system comprises three coupled nonlinear differential equations, showing sensitivity to initial conditions and the presence of invariant structures like chaotic attractors.

Transient chaos is a notable phenomenon in chaos theory, characterized by systems temporarily exhibiting chaotic behavior before settling into a final state. In many cases, transient chaos can be beneficial. For instance, chaotic vibrations can enhance the efficiency of energy harvesters [3]. Similarly, chaotic dynamics in ecological models can stabilize populations and prevent collapse [4]. To address the challenge of managing transient chaos, researchers have developed techniques like partial control, which rely on safe sets and safety functions [4], to confine chaotic trajectories within desired regions of phase space [5, 6].

The safety function is an essential tool for confining chaotic transients in systems described by difference equations. By determining the minimal control needed to keep trajectories within a specified phase-space region over a certain number of iterations [7, 8], it provides a robust method for chaos control. While controlling noise-free systems is relatively simple, achieving low control efforts in noisy scenarios presents a greater challenge. However, computing the safety function is resource-intensive, particularly for high-dimensional systems or real-time applications. This computational complexity limits its broader application in scenarios where rapid solutions are critical.

Recent advances in machine learning have introduced efficient alternatives for estimating safety functions [9]. Transformer-based models, in particular, have shown great promise in estimating the convergence behavior of the safety function for a wide range of dynamical systems without the need for fine-tuning. These models significantly improve computational efficiency and bypass the requirement of a physical model by relying solely on samples of trajectories from the dynamical system.

[a] e-mail: david.valle@urjc.es
[b] e-mail: ruben.capeans@urjc.es
[c] e-mail: alexandre.wagemakers@urjc.es
[d] e-mail: miguel.sanjuan@urjc.es (corresponding author)

In this paper, we explore the application of the transformer-based model proposed in [9] for estimating safety functions and safe sets in the one-dimensional Lorenz system, a simplified yet representative model of the three-dimensional Lorenz system. This approach leverages machine learning to provide accurate predictions without requiring extensive system knowledge or iterative computations, offering a novel perspective on controlling chaotic systems. By comparing the machine learning-based safety functions with those obtained from classical methods, we demonstrate the potential of this approach to enhance control strategies and provide new insights into transient chaos management.

The paper is organized as follows: Sect. 2 provides an overview of the Lorenz system, detailing its dynamics and the construction of the one-dimensional map used for control. Section 3 outlines the computation of safety functions using both classical and machine learning-based methods. Section 4 presents the results, including a detailed comparison of the safety functions and their derived safe sets, control efforts, and trajectory behaviors for both approaches. Finally, Sect. 5 discusses the key findings, limitations of the study, and potential directions for future research.

## 2 The Lorenz system

The Lorenz system, introduced by Edward Lorenz in 1963 [2], is a canonical example of chaotic dynamics in nonlinear systems. It consists of three coupled nonlinear differential equations:

$$\begin{aligned}\dot{x} &= \sigma(y - x), \\ \dot{y} &= x(r - z) - y, \\ \dot{x} &= xy - bz.\end{aligned} \quad (1)$$

Here, the dot notation (e.g., $\dot{x}$) indicates differentiation with respect to time. Depending on $\sigma$, $r$, and $b$ values, the system can exhibit periodic solutions, chaotic attractors, or transient chaos. For this study, we set $\sigma = 10$, $b = 8/3$, and $r = 20$, a regime known to produce transient chaos [10, 11].

For this choice of parameters, transient chaos in the Lorenz system is characterized by chaotic trajectories that eventually decay into one of two fixed-point attractors, denoted as $C^+$ and $C^-$. Figure 1 illustrates this behavior, where the green lines represent transient chaotic orbits that eventually escape into either the red or blue fixed-point attractors. Our goal is to utilize the safety function to prevent trajectories from settling into attractors, ensuring they remain within the transient chaotic regime.

Regarding the integration of the orbits, we performed numerical simulations of the Lorenz system using a fourth-order Runge–Kutta integrator. Initial conditions for 2000 trajectories were randomly sampled within the range $x$, $y$, $z \in [-30, 30]$, with an integration step size of $h = 0.001$ over a total duration of $t = 50$ units of time.

To incorporate noise effects, we added a stochastic term $\chi_h$ at each integration step. The noise term was modeled as $\chi_h = \eta \cdot \sqrt{h} \cdot N(0, 1)$, where $\eta$ is the noise intensity parameter, $h$ is the integration time step, and $N(0, 1)$

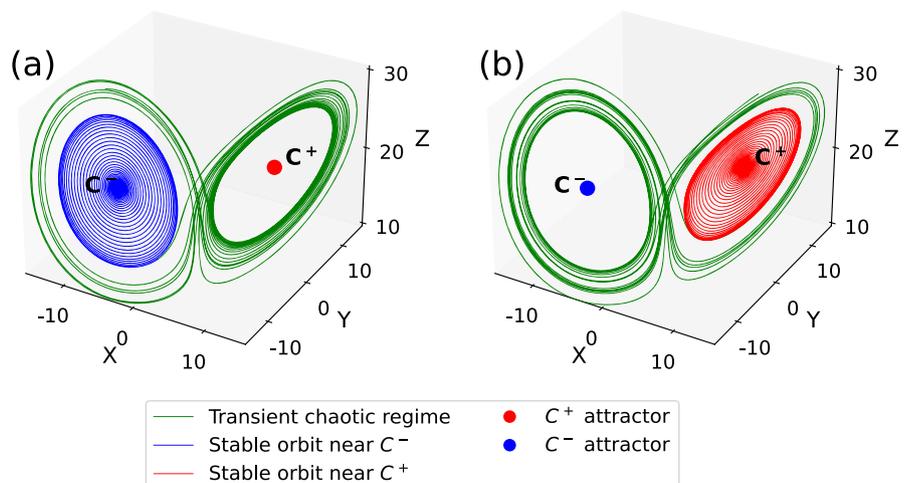

**Fig. 1** Illustration of transient chaos and the convergence to fixed-point attractors in the Lorenz system for $\sigma = 10.0$, $b = 2.67$, and $r = 20.0$. Panel (**a**) highlights the transient chaotic regime transitioning to the stable orbit near $C^-$ (blue), while panel (**b**) shows the transition to the stable orbit near $C^+$ (red). Green curves indicate the transient chaotic trajectories

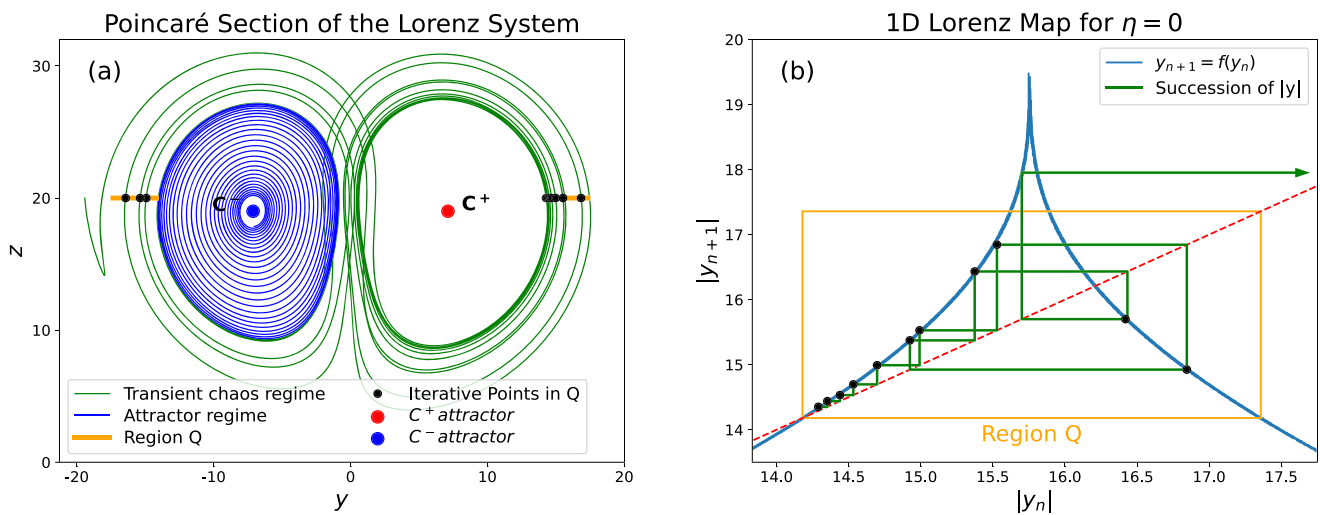

**Fig. 2** Analysis of transient chaos and iterative dynamics in the Lorenz system. (**a**) The orbit illustrates transient chaos (green line) before transitioning to the attractor regime of $C^-$ (blue line), with region $Q = 14.18 \leq |y| \leq 17.35$ (orange) marked as the target region for control. Black dots represent iterative values of $|y|$ in $Q$. (**b**) The 1D Lorenz map shows $y_{n+1} = f(y_n)$ for the case $\eta = 0$, illustrating the iterative progression and escape dynamics within $Q$

represents a random number from a standard normal distribution. Two noise intensity values, $\eta = 0.01$ and $\eta = 0.01$, were considered in our simulations.

To analyze and control the dynamics, we apply a Poincaré section defined by the conditions $x \in [-30, 30]$, $12 \leq |y| \leq 22$, and $z = 20$. This choice defines two separated sections in the phase space, one on the left and one on the right as shown in Fig. 2a.

Within this section, we focus on a target region $Q$, defined as $14.18 \leq |y| \leq 17.35$, where we aim for trajectories to be confined. By examining the successive values of $|y|$ within the Poincaré section, we construct a one-dimensional (1D) map that simplifies the three-dimensional (3D) dynamics into a computationally efficient framework for studying and controlling chaotic trajectories.

The reduced 1D Lorenz map provides a clear relationship of the form $y_{n+1} = f(y_n)$, linking successive values of $|y|$. This mapping is crucial for implementing the safety function, as it allows us to estimate the control required to keep trajectories within the region $Q$. Figure 2b demonstrates how trajectories behave within this framework, emphasizing the iterative dynamics and their eventual escape when control is not applied. By maintaining the trajectories within $Q$, we aim to extend the chaotic transient regime indefinitely.

As mentioned before, we considered two cases of study in the Lorenz system: $\eta = 0.01$ and $\eta = 0.1$, these noisy orbits propagate to the sampled data in the 1D Lorenz map. These deviations obscure the true functional relationship $y_{n+1} = f(y_n)$. To identify the underlying function $f$ and quantify the bounds of noise, we employ *conditional quantile estimation using kernel smoothing* in MATLAB [12]. This method generates three key outputs: a median curve representing the central trend of $f(y_n)$, which indicates the most likely relationship between successive $|y|$ values; and upper and lower quantile curves, which define the noise-induced bounds, capturing the maximum deviations in the sampled data.

The process begins by splitting the noisy samples into left and right branches, followed by training random forest models for each branch. Smoothed curves are then generated using the *locally estimated scatterplot smoothing* (LOESS) regression method to provide a detailed representation of the data. In our 1D Lorenz map, the noise distribution is neither uniform nor consistent across all points, adding complexity to the modeling process. To estimate the noise bounds at each point in $f(y_n)$, we sampled 2000 points from various orbits crossing $Q$ in the Poincaré section. This approach offers a robust framework for accurately modeling $f(y_n)$ and capturing noise-related uncertainties. These steps are critical for deriving an approximate model of the 1D Lorenz equations. With this model established, we now shift our focus to controlling the transient trajectory.

## 3 Computation of safety functions

Partial control is a powerful technique for controlling chaotic systems. It focuses on confining trajectories within a specific region of phase space by applying minimal control. This method leverages a concept called the safe set, a subset of the phase space where chaotic trajectories can be confined indefinitely despite the presence of

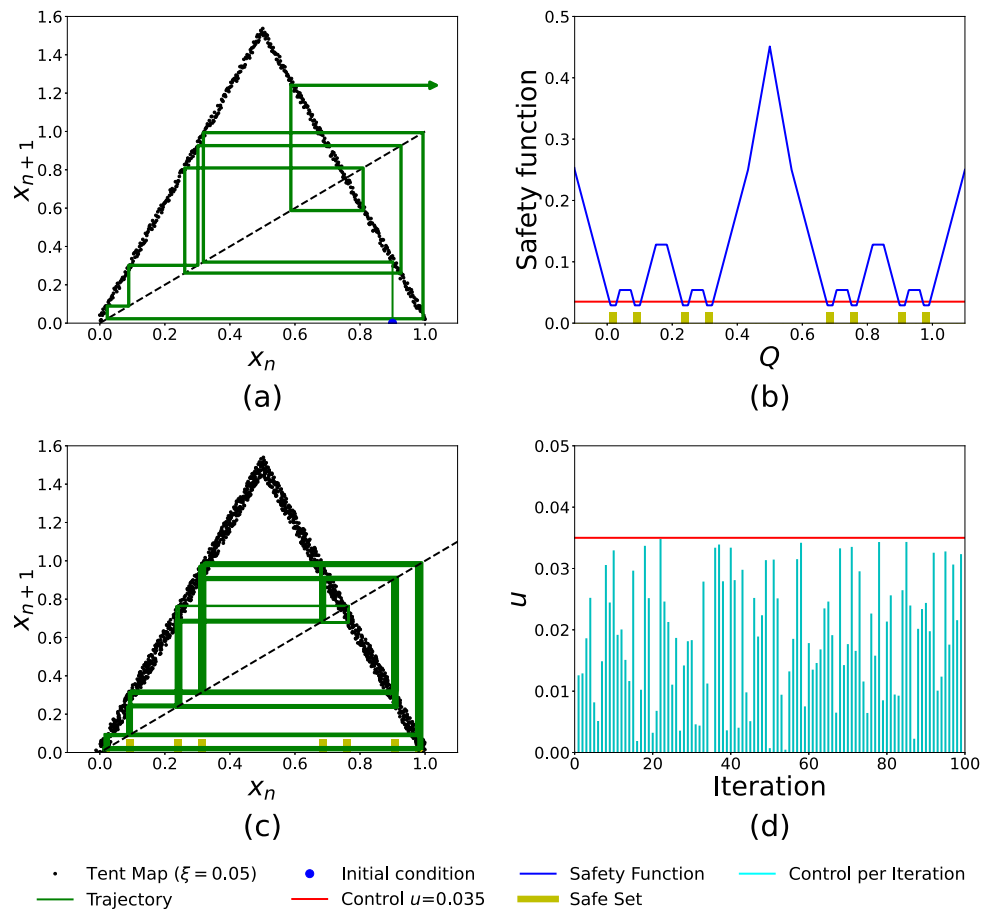

**Fig. 3** Computation of safety function for the tent map under noise $\xi = 0.05$.
(**a**) Uncontrolled trajectory escaping $Q = [0, 1]$.
(**b**) Computed safety function $U(x)$ showing the safe set corresponding to $U(x) \leq u = 0.035$.
(**c**) Controlled trajectory confined to the safe set.
(**d**) Control values applied over 100 iterations disturbances. The safe set exists within a region $Q$, which contains a chaotic saddle, and its boundaries are determined by the interplay between the system's noise level and the available control.

Safety functions are instrumental in deriving and utilizing safe sets. They quantify the minimum control required to prevent a trajectory from leaving $Q$. By identifying points within $Q$ where the control is below a specific threshold, the safe set can be derived. This approach enables the partial control method to keep trajectories into the chaotic regime, avoiding collapse into a final state. To illustrate this, consider the slope-3 tent map:

$$x_{n+1} = \begin{cases} 3x_n + \xi_n + u_n & \text{if } x_n \leq 0.5, \\ 3(1 - x_n) + \xi_n + u_n & \text{if } x_n > 0.5. \end{cases} \quad (2)$$

This map exhibits transient chaos within the interval $Q = [0, 1]$. Here, noise $\xi_n \leq \xi$ and control $u_n \leq u$ influence the dynamics. Figure 3 illustrates the computation of the safety function for this map under noise $\xi = 0.05$. Panel (a) shows an uncontrolled trajectory escaping $Q$, while panel (b) depicts the safety function $U(x)$, and a control value $u = 0.035$ which defines the safe set as the values in $Q$ where $U(x) \leq u$. Trajectories in the safe set remain in a transient orbit as shown in panel (c). Finally panel (d) highlights the control values applied during the first 100 iterations, demonstrating the effectiveness of the safety function.

### 3.1 Classical computation of safety functions

In the classical approach, the region $Q$ is discretized into $N$ grid points, where $N$ denotes the total number of spatial points used for the approximation of $Q$. For each grid point $q_i \in Q$, the map $f(q_i, \xi_s)$ is evaluated across all possible disturbance scenarios $\xi_s$, where $s$ indexes a set of $M$ representative disturbances. This approach explicitly accounts for noise by generating a range of possible outcomes, or "disturbed images", for each $q_i$. These images capture how the system can evolve under different noise levels, illustrating all potential next states.

The safety function is computed iteratively, using these disturbed bounds and prior safety function values to determine the control required to stabilize chaotic trajectories under noise. This process continues until convergence, yielding $U_\infty[q_i]$, the minimum control necessary to confine trajectories within the safe region $Q$. The iterative

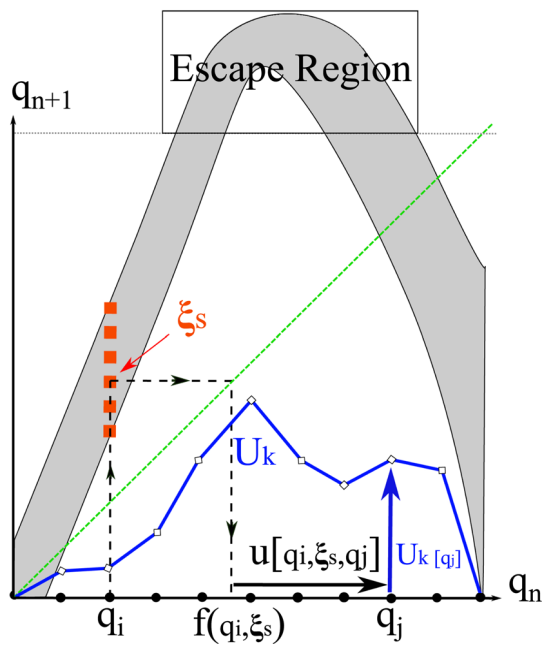

**Fig. 4** Scheme of a map affected by a bounded disturbance distribution. To compute the function $U_{k+1}[q_i]$, we have to consider all disturbed images $f(q_i, \xi_s)$ corresponding to the point $q_i$. Then compute all the corresponding control bounds, and finally extract the maximum control among them all computation is summarized in Eq. (3) and illustrated in Fig. 4.

$$U_{k+1}[q_i] = \max_{1 \leq s \leq M} \left( \min_{1 \leq j \leq N} \left( \max \left( u[q_i, \xi_s, q_j], U_k[q_j] \right) \right) \right). \tag{3}$$

In this equation, the outer maximization evaluates the worst-case scenario across all disturbances $\xi_s$, ensuring the safety function accounts for the most challenging conditions. The minimization calculates the least control required to transition from $f(q_i, \xi_s)$ to any possible next state $q_j$, while the inner maximization ensures the control required for this transition is compared to $U_k[q_j]$ from the previous iteration. This iterative process ensures convergence to $U_\infty[q_i]$, confining trajectories indefinitely within $Q$ under minimal control.

By systematically evaluating all potential disturbances and transitions, the classical approach provides a robust framework for computing safety functions and identifying safe sets. This method ensures that chaotic trajectories remain stable under noisy conditions. However, its computation involves three nested minimization and maximization processes, which make it computationally intensive and challenging for higher-dimensional systems. To address this limitation, the machine learning-based computation of safety functions was developed as a more efficient alternative.

### 3.2 Estimation of safety functions using machine learning

Recent advances in machine learning, particularly transformer-based models, provide a data-driven alternative to the classical computation of safety functions. Unlike traditional methods, which require explicit knowledge of the system's dynamics, these models estimate $U_\infty(q)$ directly from noisy trajectory samples. By leveraging large, diverse datasets of chaotic systems, transformer models are trained to identify patterns and relationships that allow them to predict safety functions accurately.

The primary advantage of this approach lies in its flexibility and computational performance. Transformer-based models can generalize across different systems, making them particularly useful for scenarios where analytical models are unavailable or computationally prohibitive. Once trained, these models bypass the need for iterative computations, producing near-instantaneous predictions of the safety function for new trajectory samples.

The machine learning estimation uses a transformer-based neural network previously described in [9]. This model includes two transformer blocks, two convolutional blocks, pooling layers, and dense layers, totaling about 1.499.661 trainable parameters. The network was trained on orbits from pseudorandom one-dimensional systems with uniformly distributed noise. Datasets of 2000 points were used, as this sample size provided sufficient information to accurately estimate the machine learning safety function. In total, roughly 26 million samples were processed over 500 epochs, ensuring robust performance.

Preprocessing was crucial to align the input data with the training conditions. The trajectory data from the Lorenz system was normalized such that the $x$-axis values ranged between [0, 1] and the $y$-axis values between [0, 1.5]. These normalization steps ensured compatibility with the network's architecture and improved prediction accuracy. Following prediction, an inverse transformation was applied to restore the results to the original scale of the Lorenz system.

For further details on the training process and architectural design of the transformer-based model, readers are encouraged to consult [9, 13].

## 4 Results and comparisons

In this section, we compare the classical and machine learning-based methods for computing safety functions and analyze their effectiveness in controlling chaotic trajectories in the 1D Lorenz system.

The safety functions are analyzed for two scenarios: in the first one, the 1D Lorenz map is derived from a Poincaré section sampling orbits of the 3D Lorenz system under noise $\eta = 0.01$, and in the second one, the orbits experience a higher noise level of $\eta = 0.1$. In both cases, we sampled 2000 points from $Q$. These samples were used for the *conditional quantile estimation using kernel smoothing* to estimate $f(y_n)$ and $\xi_n$ in the classical safety function, as well as input data for the transformer-based neural network to predict the machine learning safety function.

Figure 5 presents the computed safety functions for both the classical and machine learning approaches under these conditions. Below each plot, a line highlights the regions of divergence between the two methods: red segments mark where the machine learning function predicts higher values than the classical method, while blue segments indicate the opposite.

After obtaining the safety functions for the two cases, we evaluated their effectiveness in computing safe sets and controlling a trajectory using the control values suggested by the safety function over 100 iterations.

In Fig. 6, we illustrate the behavior of a controlled trajectory starting at the initial condition $y_0 = 0.26$ in the 1D Lorenz map constructed from 3D orbits with $\eta = 0.01$. Panels (a), (b), and (c) depict the results using the classically computed safety function, while panels (d), (e), and (f) showcase the results from the machine learning-based safety function.

Panels (a) and (d) show the trajectories for the same initial condition, each following different control values during iteration, along with their respective safe sets, defined as all points in $Q$ where $U_\infty(q) \leq U_\infty(y_0)$. While the trajectories appear similar, discrepancies in the safe sets are evident: some regions align, while others are either missing or overrepresented. These differences are highlighted in panel (d) with red regions indicating overestimations and blue regions representing underestimations compared to the classical safe set.

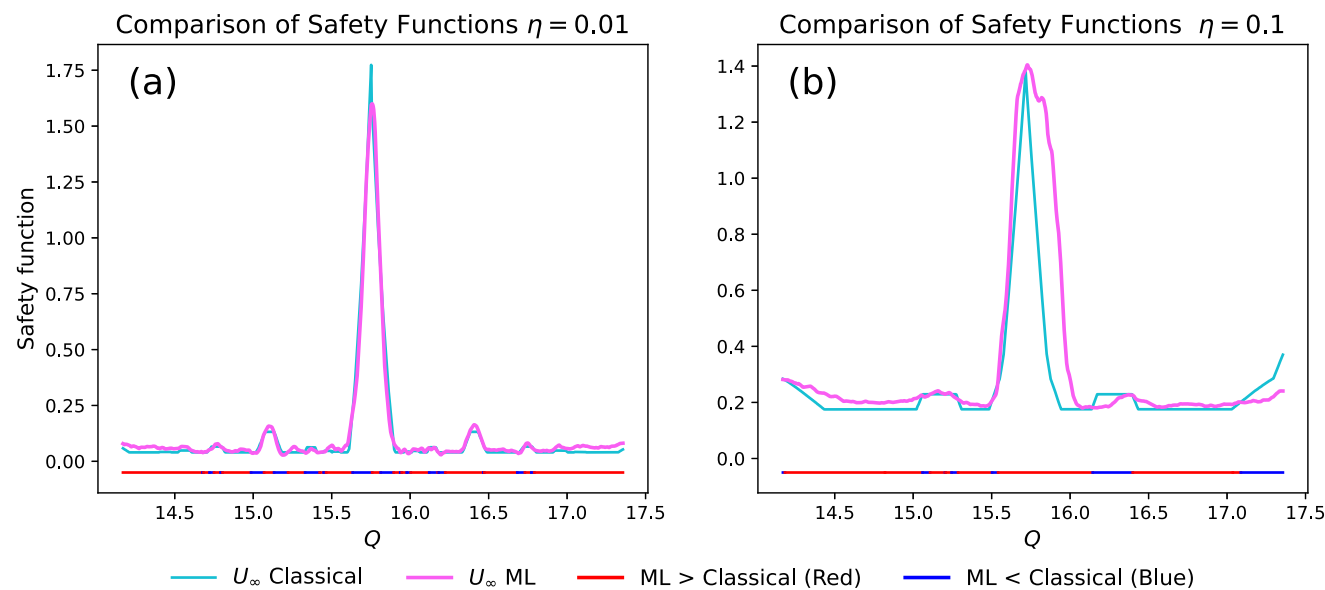

**Fig. 5** Comparison of safety functions for the 1D Lorenz map under two noise levels. (**a**) Safety functions computed for $\eta = 0.01$. (**b**) Safety functions computed for $\eta = 0.1$. In both panels, the classical safety function ($U_\infty$ Classical) is shown in cyan, while the machine learning-based safety function ($U_\infty$ ML) is shown in magenta. The lines below each plot indicate regions of disagreement: red segments denote where the ML-based function exceeds the classical function, and blue segments highlight where the classical function exceeds the ML-based function

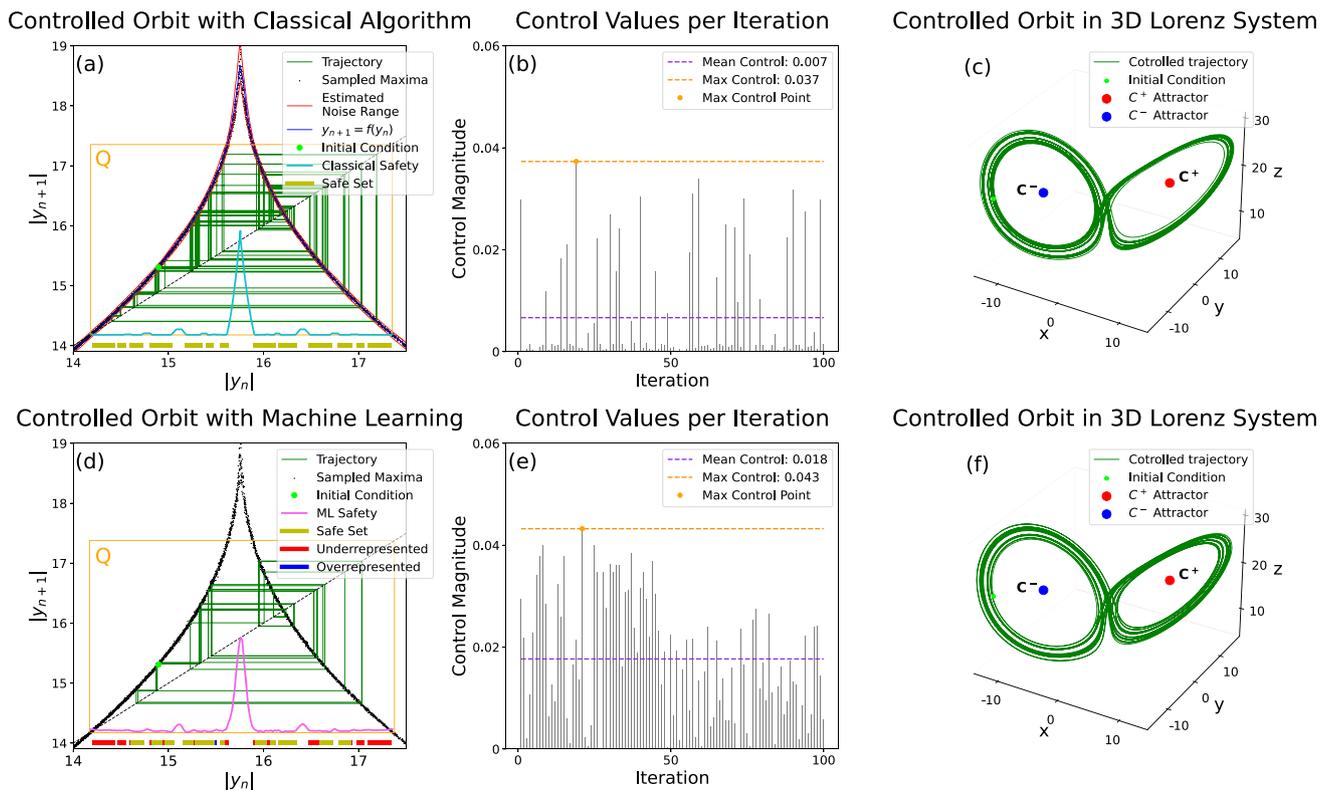

**Fig. 6** Comparison of controlled orbits using classical and machine learning safety functions for the case $\eta = 0.01$. (**a**) Safe set, trajectory, and sampled maxima for the classical safety function, with control values shown in (**b**) and the corresponding 3D Lorenz orbit in (**c**). (**d**) Safe set, trajectory, and maxima for the machine learning-based safety function, highlighting overestimated (red) and underestimated (blue) regions compared to the classical safe set. (**e**) Control values for the machine learning approach, with the corresponding 3D Lorenz orbit in (**f**). Both methods successfully confine the trajectory to the transient chaotic regime, preventing its convergence to the fixed-point attractors

Panels (b) and (e) present the control values applied during each iteration of the trajectory. The machine learning-based approach, while effective, tends to produce higher control magnitudes compared to the classical method. This highlights a trade-off between the simplicity of the machine learning model and the precision of the classical computation.

Finally, panels (c) and (f) display the corresponding 3D trajectories in the Lorenz system. Both safety functions successfully confine the trajectory within the transient chaotic region, preventing escape into the attractors. This demonstrates that, despite differences in control effort and safe set representation, both approaches effectively stabilize chaotic trajectories.

For the case where the 1D Lorenz map was derived from 3D orbits with noise $\eta = 0.1$, the control performance of both the classical and machine learning-based safety functions is shown in Fig. 7. Panels (a), (b), and (c) correspond to the classical approach, while panels (d), (e), and (f) show the machine learning results.

The trajectories and safe sets in panels (a) and (d) demonstrate that both methods confine the orbit within the transient chaotic regime. The machine learning-based safe set once again shows overestimated (red) and underestimated (blue) regions compared to the classical safe set, reflecting the impact of higher noise on the estimation.

Panels (b) and (e) reveal that the machine learning-based approach requires larger control magnitudes on average to stabilize the trajectory, as indicated by the mean control values. Despite this, the trajectory remains successfully controlled, as shown in panels (c) and (f), where both methods keep the orbit within the transient chaotic region of the 3D Lorenz system.

When comparing the safe sets for both scenarios, the new algorithm consistently produces overestimated and underestimated regions relative to the classical safety function. The underestimated regions (red) represent parts of the safe set excluded by the machine learning prediction due to overestimating the required control. In these cases, the trajectory could remain stable, but the predicted safe set unnecessarily excludes these areas. While this does not compromise stability, it leads to the application of higher than needed control values.

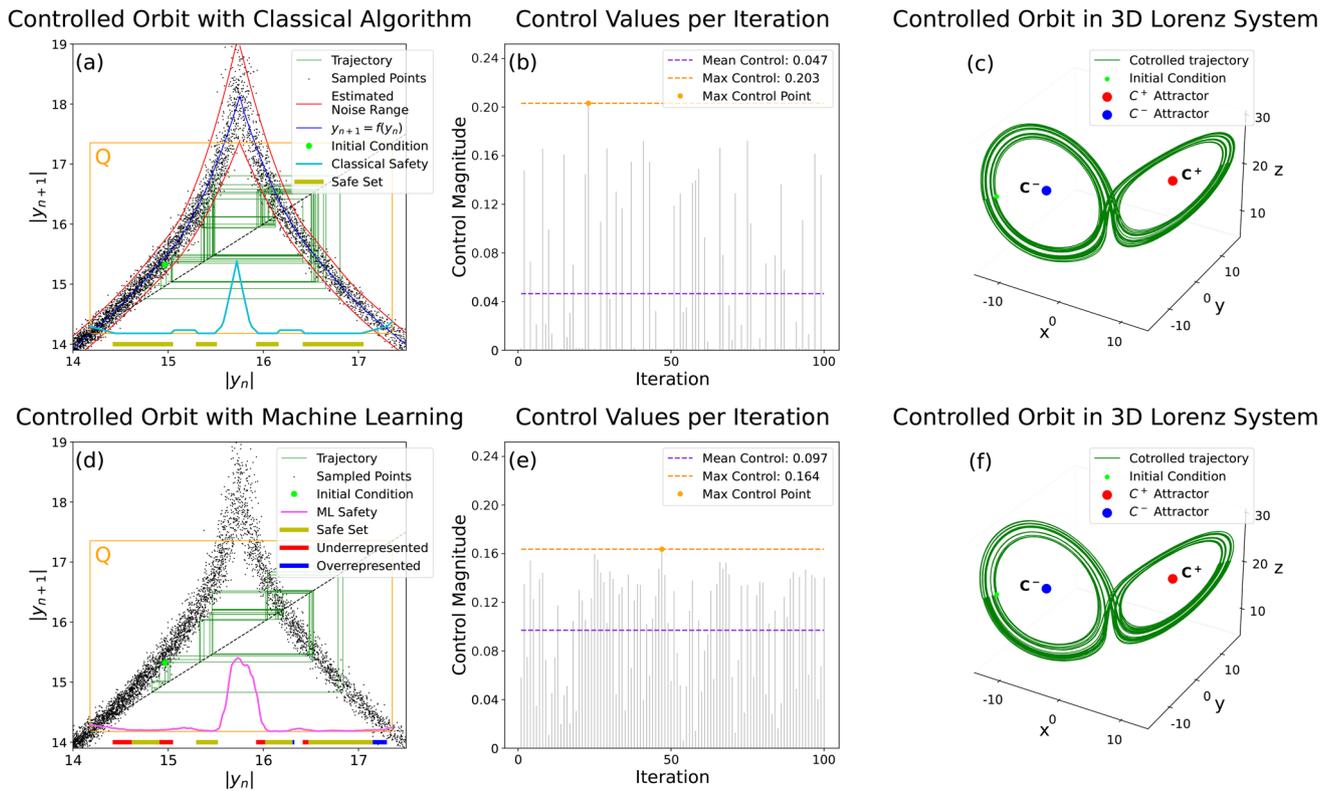

**Fig. 7** Controlled orbits under noise $\eta = 0.1$ using classical and machine learning safety functions. (**a**) Classical safe set and trajectory with applied control values shown in (**b**) and corresponding 3D Lorenz orbit in (**c**). (**d**) Machine learning safe set and trajectory, highlighting overestimated (red) and underestimated (blue) regions, with control values in (**e**) and the 3D Lorenz orbit in (**f**). Both methods maintain the trajectory in the transient chaotic regime

Conversely, overestimated regions (blue) occur where the machine learning algorithm underestimates the control required, predicting stability in areas where the trajectory is more likely to escape. If an initial condition lies within these regions, the orbit may leave the predicted safe set unless additional control is applied. These discrepancies can create a feedback loop, causing inaccuracies in the predicted trajectory and applied control. Although the machine learning approach demonstrates the capability to maintain stability overall, these localized errors highlight areas where the algorithm's predictions can be refined to more closely align with the classical method.

## 5 Conclusion and future work

This paper demonstrates the effectiveness of a machine-learning approach for controlling transient chaos in the Lorenz system via safety functions derived from the simplified 1D Lorenz map. By designing a transformer-based model, we have obtained consistent predictions of the safety function and accurately captured the qualitative structure of the safe set. In particular, although the neural network was trained on uniformly distributed noise, it still has performed well when faced with the non-uniform noise distribution of the 1D Lorenz system. Thus, the predicted safety function successfully enables chaos control under noisy conditions, making it a viable alternative to classical methods.

A key observation from the results is the presence of overestimated and underestimated regions in the predicted safe sets compared to those derived from classical computation. Overestimated regions provide a conservative buffer that ensures trajectory stability but may lead to higher-than-necessary control values. Conversely, underestimated regions exclude parts of the safe set where the stability could still be maintained, potentially introducing inaccuracies in the predicted control effort. Refining the model to address these discrepancies, such as through uncertainty quantification, could significantly enhance its reliability and performance.

Future work will aim to expand the applicability and robustness of the machine learning safety function estimation. A critical objective will be to extend the methodology to higher-dimensional chaotic systems, which present more complex dynamics. Such extensions will provide an opportunity to demonstrate the scalability of the approach and its potential to enhance the efficiency of the safety function computation and real-time control decision-making.

Overall, the proposed approach represents a promising advancement in chaos control. Its ability to maintain stability under varying noise levels and its scalability to more complex systems highlight its potential as a robust and efficient tool for analyzing and controlling chaotic systems in both theoretical and practical contexts.

**Acknowledgements** This work has been financially supported by the Spanish State Research Agency (AEI) and the European Regional Development Fund (ERDF, EU) under Project No. PID2023-148160NB-I00 (MCIN/AEI/10.13039/501100011033).

**Funding** Open Access funding provided thanks to the CRUE-CSIC agreement with Springer Nature.

**Data Availability** The authors declare that the data supporting the findings of this work are available within the following repository [13].